\documentclass[aip,graphicx]{revtex4-1}
\usepackage{graphicx}
\draft % marks overfull lines with a black rule on the right

\begin{document}

%\preprint{}

\title{Cosmic ray muon computed tomography of spent nuclear fuel in dry storage casks} %Title of paper

\author{D. Poulson}
\email[Corresponding author.  Email address: ]{$poulsond@lanl.gov$}
\affiliation{Los Alamos National Laboratory, Los Alamos, NM 87544 USA}
\affiliation{University of New Mexico, Albuquerque, NM 87131 USA}

\author{J. M. Durham}
\affiliation{Los Alamos National Laboratory, Los Alamos, NM 87544 USA}

\author{E. Guardincerri}
\affiliation{Los Alamos National Laboratory, Los Alamos, NM 87544 USA}

\author{C. L. Morris}
\affiliation{Los Alamos National Laboratory, Los Alamos, NM 87544 USA}

\author{J. D. Bacon}
\affiliation{Los Alamos National Laboratory, Los Alamos, NM 87544 USA}

\author{K. Plaud-Ramos}
\affiliation{Los Alamos National Laboratory, Los Alamos, NM 87544 USA}

\author{D. Morley}
\affiliation{Los Alamos National Laboratory, Los Alamos, NM 87544 USA}

\author{A. Hecht}
\affiliation{University of New Mexico, Albuquerque, NM 87131 USA}

\date{\today}

\begin{abstract}
Radiography with cosmic ray muon scattering has proven to be a successful method of imaging nuclear material through heavy shielding.  Of particular interest is monitoring dry storage casks for diversion of plutonium contained in spent reactor fuel.  Using muon tracking detectors that surround a cylindrical cask, cosmic ray muon scattering can be simultaneously measured from all azimuthal angles, giving complete tomographic coverage of the cask interior. This paper describes the first application of filtered back projection algorithms, typically used in medical imaging, to cosmic ray muon imaging.  The specific application to monitoring spent nuclear fuel in dry storage casks is investigated via GEANT4 simulations.  With a cylindrical muon tracking detector surrounding a typical spent fuel cask, the cask contents can be confirmed with high confidence in less than two days exposure.  Similar results can be obtained by moving a smaller detector to view the cask from multiple angles.
\end{abstract}


\maketitle %\maketitle must follow title, authors, abstract and \pacs
LA-UR-16-21971
\section{Introduction}

The method of reconstructing multidimensional functions from a continuum of integrated projections at varying angles was first discussed by Radon \cite{Radon}. 	Applications of this technique to produce tomographic images of the interior of the human body via series of x-ray images were first formalized by Cormack \cite{Cormack1,Cormack2} and demonstrated by Hounsfield \cite{Hounsfield} and Ambrose \cite{Ambrose}. These computed tomography techniques were soon extended to charged particle imaging with alpha particles \cite{Goitein} and heavy nuclei \cite{HI_rad} produced by particle accelerators.  Here, we apply computed tomography to imaging with a natural source of charged particles: cosmic ray muons.  Through simulation, we explore the technique for imaging spent nuclear fuel in dry storage casks.

Cosmic ray muons are produced by interactions of protons and nuclei from space with atoms in the upper atmosphere.  Collisions of these primary cosmic rays with atmospheric gas produce showers of pions, many of which decay to muons.  These muons arrive on the Earth’s surface at a rate of ~1/cm$^{2}$/min with an average energy of ~4 GeV, and a flux that falls off approximately with $cos^{2}\theta_{z}$, where $\theta_{z}$ is the angle from the zenith \cite{PDG}.   Since the flux of muons is isotropic in azimuth, they form a naturally occurring probe that can be used to radiograph objects from multiple views simultaneously.

Due to their high energies, ubiquitous availability, and highly penetrating nature, comic ray muons are being explored as a tool for a variety of difficult imaging scenarios \cite{MURAY, Nagamine1, Hungarians, FukuPRL, MorrisTCNA, SugitaTCNA}.  Of particular interest to international safeguards is monitoring spent nuclear fuel that is sealed in dry storage casks to deter potential diversion to plutonium reprocessing.  These casks are heavily shielded to prevent radiation leakage to the environment, which precludes quantitative imaging using typical radiographic probes such as neutrons or photons \cite{Ziock}.  Other methods that attempt to monitor casks by obtaining a neutron "fingerprint" can only potentially determine if the content of the cask has changed since it was first measured \cite{NENstuff1}, and are therefore only applicable to casks which were measured when the contents were definitively known by the inspection agency.  Also, these measurements must apply corrections for changes in the neutron emission rates due to decay of fission products between measurements, and are susceptible to large backgrounds from neighboring casks that are present at spent fuel storage installations (SFSIs).  Muon radiography is the only standalone method which can determine a cask's contents without any prior information and allow inspectors to obtain or recover from a loss of continuity of knowledge.  

Simulation studies to date have used two planar tracking detectors placed on opposite sides of an object to record the trajectories of incoming and outgoing muons, and thereby determine the scattering angle \cite{SwedenThesis, ClarksonSim, Jonk, AmbrosinoSim, SteliosTNS, SteliosNIM}.  The first actual muon radiography measurements on a fuel cask used two 1.2 x 1.2 m$^{2}$ drift tube tracking detectors to measure the areal density between the two detectors, and thereby show that missing fuel assemblies can be located \cite{DurhamCask}.  However, this single measurement of integrated density leaves ambiguities in the exact location of the missing fuel elements in the direction orthogonal to the two detectors.  

For these large, cylindrical storage containers, a more optimal geometry is a cylindrical detector that fits around the outside of the cask.  In addition to increasing the detector active area (and thereby the muon counting rates), this provides simultaneous measurements of muon scattering through all azimuthal angles of the cask.  With this continuum of areal density measurements, computed tomography image reconstruction algorithms can be applied to produce full images of the cask interior.  A single set of planar detectors can also take data from multiple viewing angles to provide a similar coverage on a longer time scale.

This paper describes the feasibility of such a technique using GEANT4 simulations \cite{G4} of cosmic ray muons passing through a idealized cylindrical muon tracking detector around a partially loaded Westinghouse MC-10 cask \cite{MC10}.  A filtered back-projection algorithm is used to reconstruct images of the cask interior, which shows that missing assemblies and missing portions of individual fuel assemblies can be identified without opening the cask.  Similar data can also be obtained with a series of smaller muon tracking detectors that are moved around the cask. This technique could be useful for international nuclear safeguards inspectors. 

\section{Simulation}

\begin{figure}[htbp]
 \begin{minipage}{0.5\linewidth}
  \centering
 \includegraphics[width=0.8\textwidth]{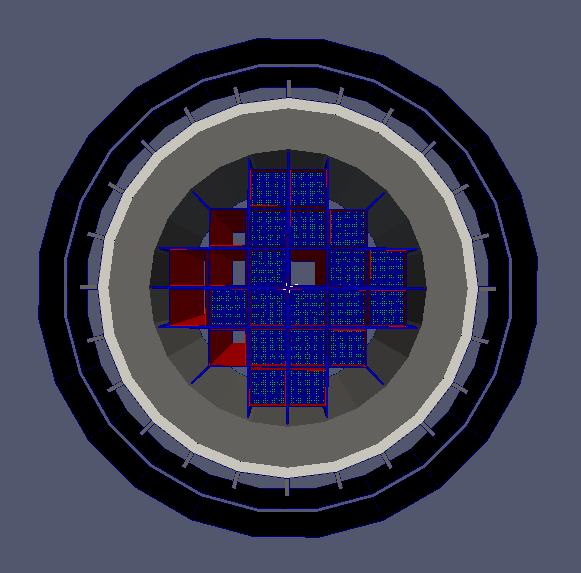}
 \end{minipage}%
 \begin{minipage}{0.5\linewidth}
  \centering
  \includegraphics[width=0.7\textwidth]{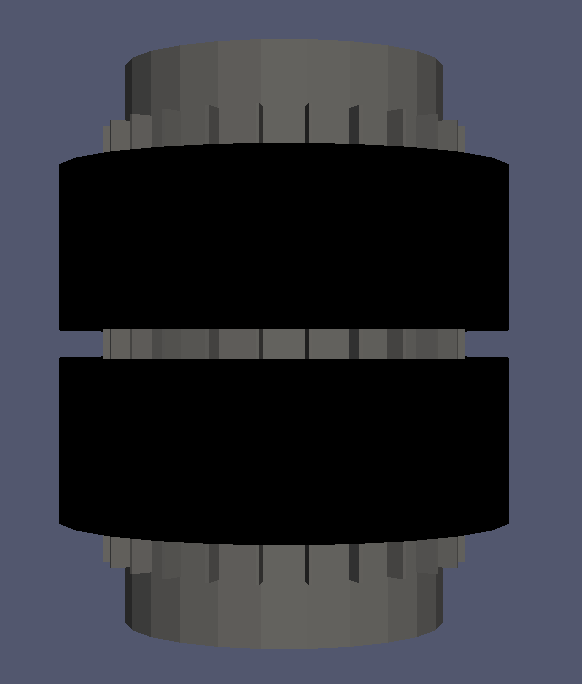}

 \end{minipage}
 \caption{A top down view (left) and a side view (right) of the simulated cask and idealized detector geometries. The fuel assembly loading configuration is shown with six fuel assembly compartments left empty.}
 \label{fig:G4casks}
\end{figure}

The GEANT4 toolkit was used to simulate muon passage through a partially loaded dry storage cask.  The simulated cask and fuel assemblies were modeled after the specifications for a Westinghouse MC-10 spent fuel cask located at Idaho National Lab.  The MC-10 is a vertical storage cask designed to hold 24 pressurized water reactor fuel assemblies.  In the simulation, the cask is partially loaded with Westinghouse 15x15 pressurized water reactor fuel assemblies in 18 of the slots (see Fig. \ref{fig:G4casks}).  An aluminum basket holds the fuel assemblies in the center of the cask, and is surrounded by a 25 cm thick forged steel wall.  This is surrounded by a shell of BISCO NS-3 neutron shielding, which is covered by a stainless steel skin.  Twenty four cooling fins are equally spaced around the cask and facilitate heat transfer from the interior of the fuel basket to the outside.  The external dimensions of the cask are 4.8 m in height by 2.7 m in diameter.

The fuel assemblies are 4 m long with a cross sectional area of 21.4 x 21.4 cm$ ^{2}$, and each contains 204 fuel rods, 20 control rod guide tubes, and one instrument thimble. The fuel rods are filled with UO$_{2}$ and surrounded by a zirconium alloy cladding. The control rods and instrument thimble were simulated an empty cladding, with no fuel present. The center-to-center distance between pins is 1.43 cm. 

The ideal muon detectors in the simulation consist of two rings which encircle the cask.  Each ring is 1.5 m high and has a radius of 1.68 m.  The upper detector is separated from the lower detector by 0.5 m, and each detector is offset either above or below the center of the cask by 0.25 m.  To be considered in this analysis, muons are required to pass through the upper detector and be pointed at the lower detector.  The detectors record the muon position and direction, and are assumed to have perfect efficiency and angular resolution.

\begin{figure}
    \includegraphics[width=\textwidth]{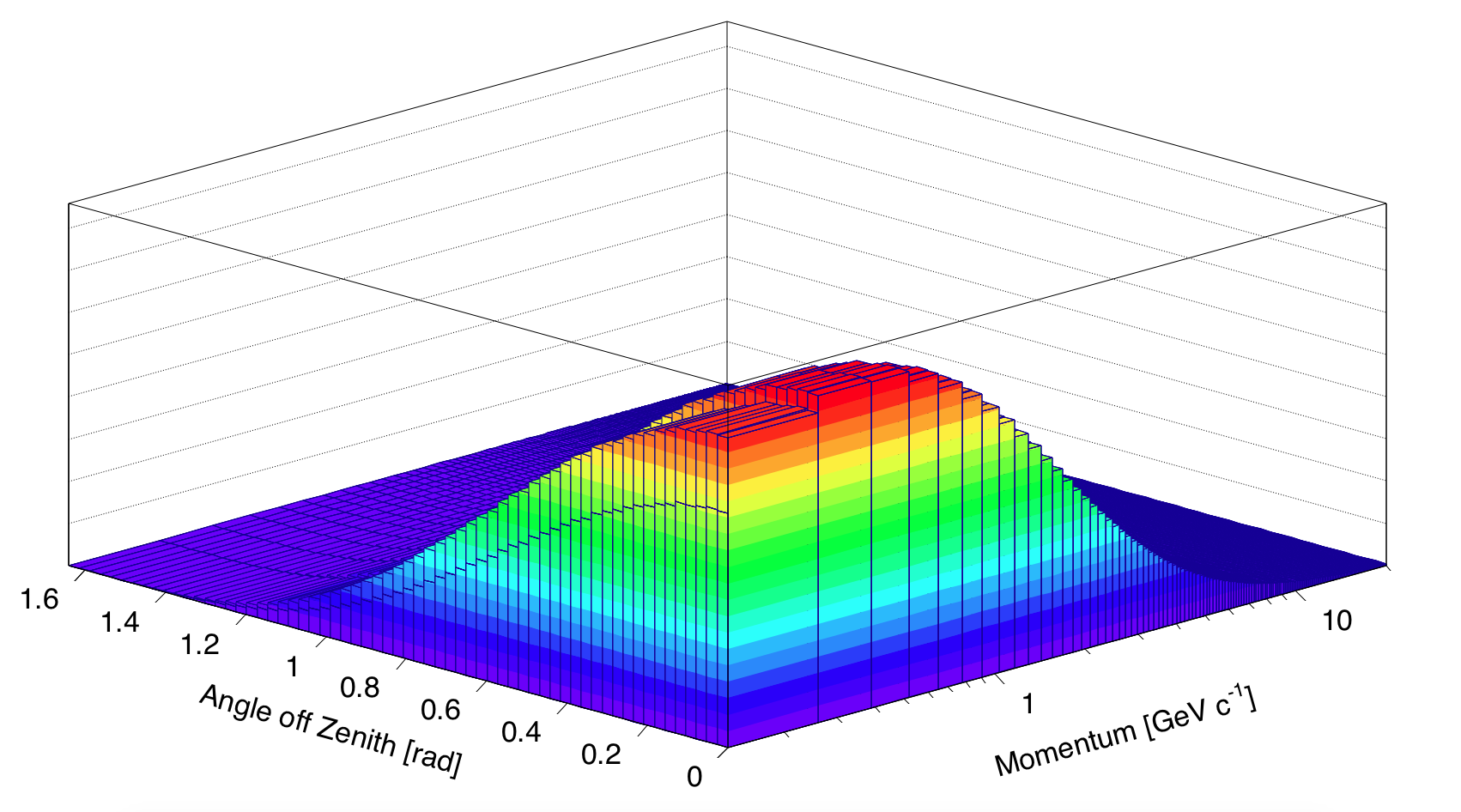}
    \caption{Monte Carlo cosmic ray muon momentum and angular distribution. Cosmic ray muons were generated in GEANT4 by sampling this distribution.}
    \label{fig:muons_in}
\end{figure}

For these studies, 10$^{8}$ muons with realistic energy and angular distributions were generated and passed through the cask. The angular distribution of the simulated muon flux follows a $cos^{2}\theta_{z}$ distribution out to close to 85 degrees from the zenith. A histogram of the muon flux as a function of muon momentum and $\theta_{z}$is shown in Fig. \ref{fig:muons_in}. An integration of the solid angle coverage of the detectors was used with the assumed muon flux of 1/cm$^{2}$/min to predict the simulated rate of recorded muons. The calculations show that the 10$^{8}$ muon events generated here are equivalent to 1.6 days of cosmic ray muon exposure.

\section{Image Analysis}

\begin{figure}
    \includegraphics[width=0.75\textwidth]{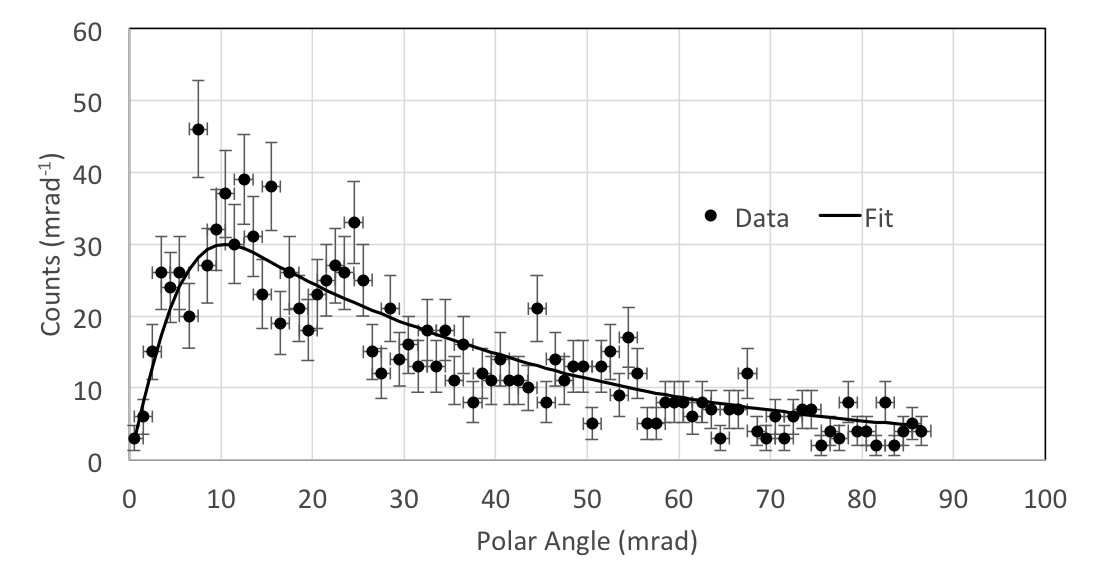}
    \caption{A typical scattering angle histogram, along with the multigroup fit.}
    \label{fig:ScatteringHistogram}
\end{figure}

Cosmic ray muons generated in the simulation impinge on the cask from all azimuthal directions. The incident muons are collected in one degree wide bins of azimuthal angle, and projected to a vertical plane in the center of the detector.  The plane is divided into 2 cm wide bins in the horizontal direction, and the scattering angle of each muon which falls within that position bin is collected in a corresponding histogram.  Once all track’s scattering angles have been assigned to a histogram, the histogram data is fit with the multigroup model \cite{multigroup} and the areal density for that bin is extracted (see Fig. \ref{fig:ScatteringHistogram}). This technique produces a series of integrated areal density projections of the cask as viewed from each azimuthal angle. 

\begin{figure}[htbp]
 \begin{minipage}{0.5\linewidth}
  \centering
  \includegraphics[width=\textwidth]{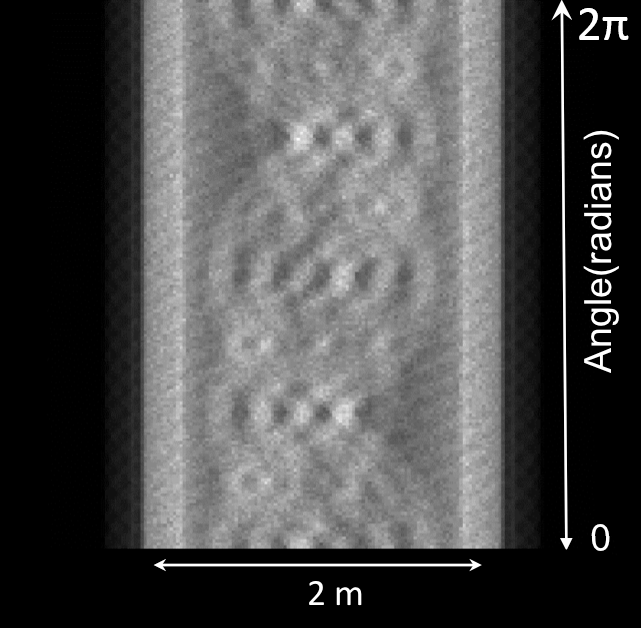}
  
 \end{minipage}%
 \begin{minipage}{0.5\linewidth}
  \centering
  \includegraphics[width=0.85\textwidth]{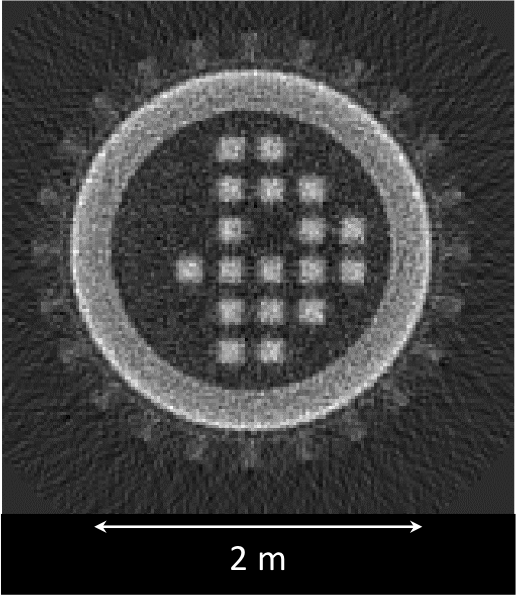}

 \end{minipage}
 \caption{A sinogram (left) and its reconstructed image via filtered back projection (right) of the areal density of the simulated cask, found with muon scattering.}
 \label{fig:sino1}
\end{figure}

Since the detectors are wrapped entirely around the cask, the complete view of the cask integrated density projections can be collected in a sinogram.  This display of the integrated areal densities versus the azimuthal projection angle is shown in the left panel of Fig. \ref{fig:sino1}.

Only the horizontal directions of the incident muon tracks determine the position bin where the muon scattering angle is stored. Therefore, the reconstructed image of areal scattering densities is a projection down the center axis of the detector. There are many tools and software packages readily available to reconstruct images from sinograms, a full discussion of which is beyond the scope of this paper. In this analysis the reconstruction of the cask image from the sinogram was performed via filtered back projection using the RECLBL library \cite{RECLBL}. The right panel of Fig. \ref{fig:sino1} shows the reconstructed image of the cask. It clearly shows the structures of the cask including the cooling fins, steel shielding shell, and the loading configuration of the fuel assemblies.

\begin{figure}[htbp]
 \begin{minipage}{0.5\linewidth}
  \centering
  \includegraphics[width=0.6\textwidth]{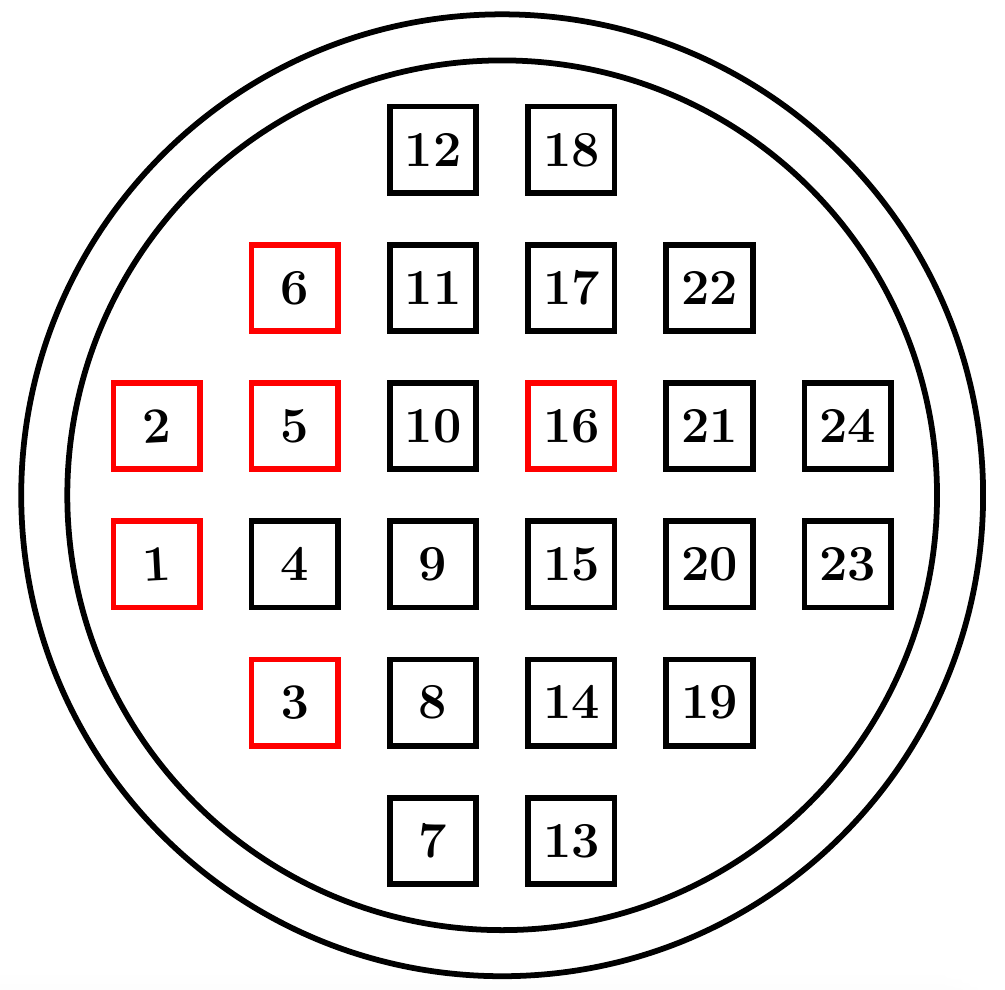}

 \end{minipage}%
 \begin{minipage}{0.5\linewidth}
  \centering
  \includegraphics[width=\textwidth]{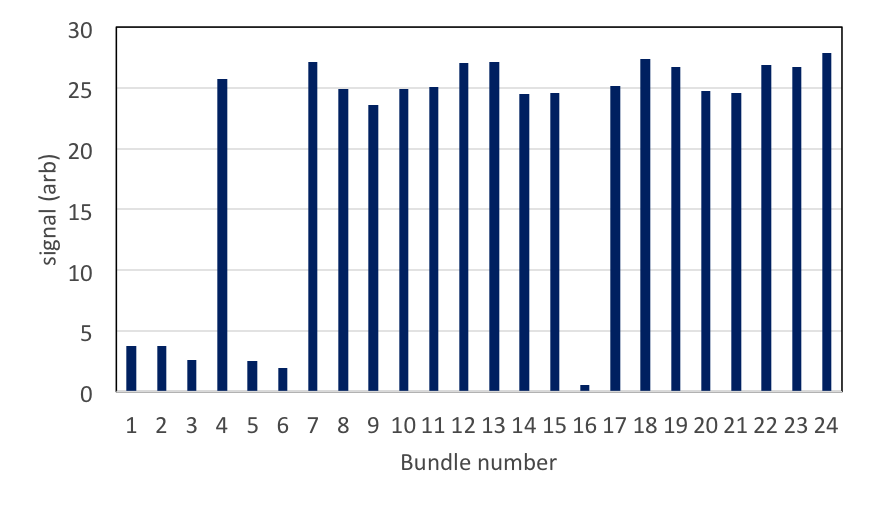}

 \end{minipage}
 \caption{The numbering scheme (left) of the fuel assemblies (red boxes indicate an empty slot) and the associated integrated signal for each slot (right). }
 \label{fig:slots}
\end{figure} 

Fig. \ref{fig:slots} shows a numbered schematic of each slot in the fuel basket along with the integrated signal at each location, which is defined as the sum of the integrated areal density values contained in a 20$\times$20 cm$^{2}$ area centered at the nominal location of the assembly. An immediate distinction is apparent between the integrated signal of slots containing the fuel assemblies and those that have been left empty.  The average integrated value of locations that contained assemblies was 25.8$\pm$1.3 arb. units. Empty slots average an integrated signal value was 2.5$\pm$1.3 arb. units. The simulation shows that 10$^{8}$ muon events, or 1.6 days, allows for sufficient exposure to distinguish between a filled and empty slot inside the cask with a statistical precision of 18$\sigma$. 

\begin{figure}[htbp]
 \begin{minipage}{0.5\linewidth}
  \centering
\includegraphics[width=1.02\textwidth]{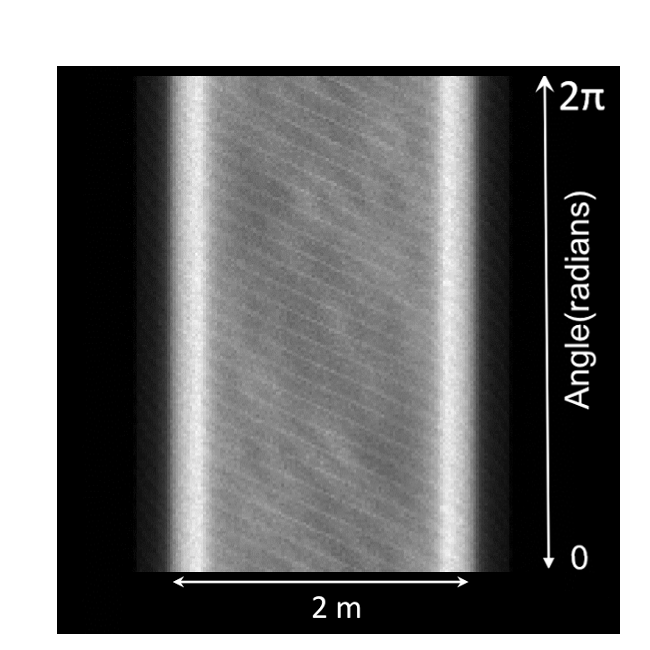}
 \end{minipage}%
 \begin{minipage}{0.5\linewidth}
  \centering
\includegraphics[width=0.8\textwidth]{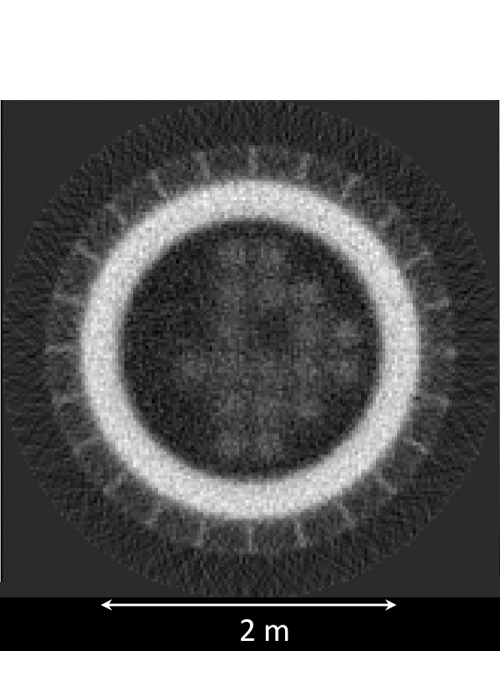}

 \end{minipage}
 \caption{A sinogram (left) and its reconstructed image via filtered back projection (right) of the muon attenuation $-ln(\frac{N_{out}}{N_{in}})$ within the cask.}
 \label{fig:sino2}
\end{figure}

We also study the transmission rate of muons through the cask.  From the simulated data set, muons which pass through the first detector but were stopped in the cask (i.e. do not reach the second detector) were selected.  Fig. 6 shows a sinogram filled using these events. The values in each bin are $-ln(\frac{N_{out}}{N_{in}})$, where $N_{out}$ is the number of tracks which pass through both the upper and lower detectors and $N_{in}$ is the total number of incoming muon tracks. The reconstructed image in Fig. \ref{fig:sino2} has high intensity in the region of the steel shell, due to stopping of low energy muons, with a lower intensity in the regions of the loaded fuel assemblies. A comparison with the reconstructed scattering image in Fig. \ref{fig:sino1} shows that the transmission image is less sensitive to the fuel bundles themselves. 

The difference between the scattering and transmission images reflects the different mechanisms governing the muon energy loss in material, which is proportional to the ratio of a materials atomic number to mass number (Z/A), and the angular distribution of scattered muons, which has a dependence on Z that is stronger than linear \cite{PDG}. Since the neutron fraction (A-Z)/A generally increases for high Z elements, the proton fraction Z/A decreases, resulting in a relatively weaker signal at the UO$_{2}$ fuel in the transmission image. Conversely, the stronger dependence on Z for multiple scattering increases the signal strength of for high Z material, providing a clearer scattering image of the fuel bundles. 

\section{Detector Geometry}

The size of the idealized muon tracking detector shown in Fig. \ref{fig:G4casks} may limit deployability at SFSIs, and the cost may be prohibitive. However, a smaller detector that collects data from multiple viewing angles around a single cask can provide tomographic information that is similar to the full idealized detector.  Such a detector was used in a previous measurement \cite{DurhamCask}, but only from one viewing angle. 

\begin{figure}[htbp]
 \begin{minipage}{0.5\linewidth}
  \centering
  \includegraphics[width=0.8\textwidth]{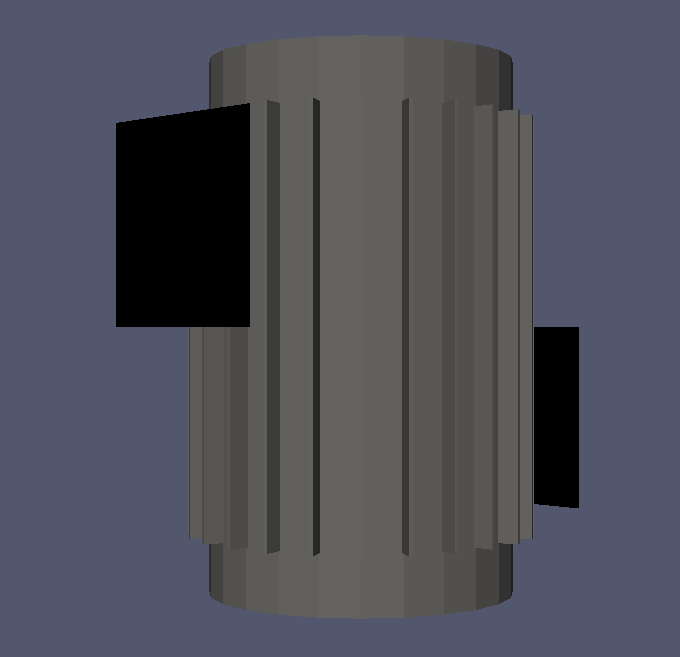}
 \end{minipage}%
 \begin{minipage}{0.5\linewidth}
  \centering
  \includegraphics[width=0.8\textwidth]{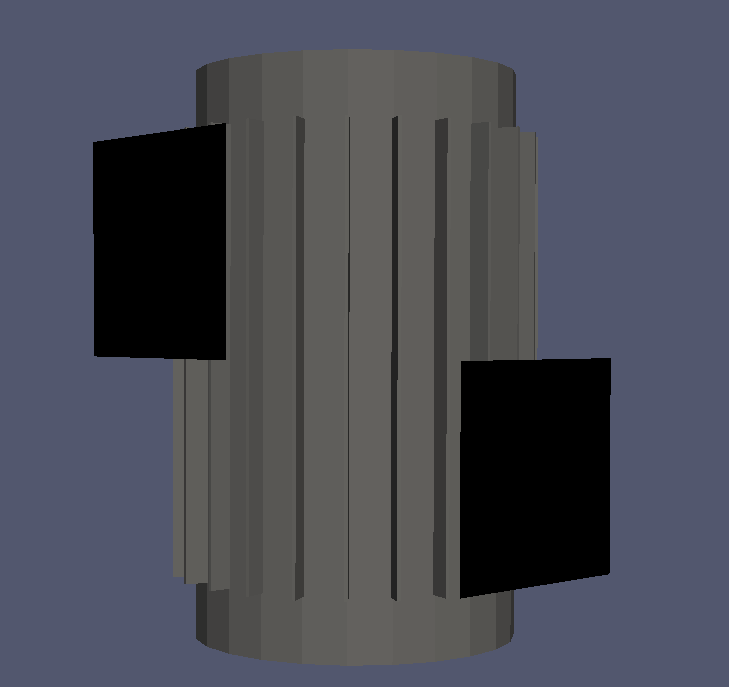}
 \end{minipage}
 \caption{Simulated small detector composed of two planar trackers to measure incident and exiting muon tracks. The lower detector on the right image is rotated by 90$^\circ$ from that of the left image to image a different volume of the cask.}
 \label{fig:MMT}
\end{figure}

With smaller upper and lower tracking detectors, the relative positions of the two planes can be changed to image different volumes of the fuel cask. Simulations were run with two 1.2$\times$1.2 m$^{2}$ trackers at two, three, and four different positions around the fuel cask (see Fig. \ref{fig:MMT}). The detectors were placed at equal angular intervals determined by the number of viewing angles. For these simulations, the upper detector was placed at one of the $n$ positions within the range (0$^\circ$,180$^\circ$] of the circumference of the cask, while the lower detector was placed at each of the $n$ positions in the range (180$^\circ$, 360$^\circ$]. Thus, the total number of data sets increased as the square of the number of positions, i.e. $n^2$ data sets for $n$ positions. 

\begin{figure}[htbp]
  \centering
  \includegraphics[width=\textwidth]{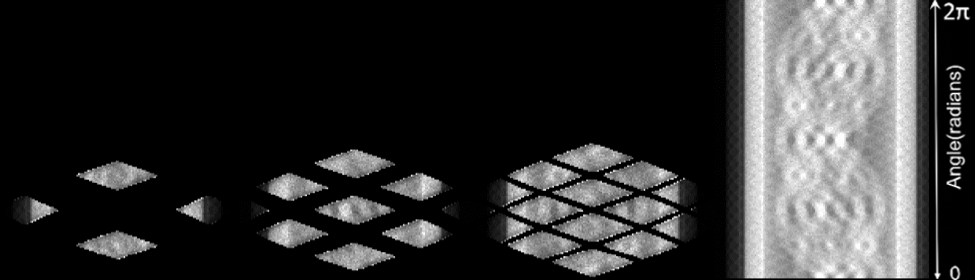} 
 \caption{Sinograms from two, three, and four small detector positions, and for the idealized cylindrical detector (left to right).}
 \label{fig:partial_sinograms}
\end{figure}

The simulated data sets for a given number of positions were combined to fill a sinogram. The resulting sinograms from each of these configurations as well as the idealized detector simulation are shown in Fig.\ref{fig:partial_sinograms}. The top half of each of the sinograms (corresponding to viewing angles $>$180$^\circ$) is empty since no muons travel up from the lower detector to the top. By increasing the number of positions, and allowing the detectors to iterate over all possible positions on both sides of the cask, the sinogram may be filled more completely. Future work will focus on image reconstruction from partial sinogram data.

\section{Technology Options}

Many different technology options for tracking charged particles are currently in use and can be adapted for cosmic ray muon tomography.  However, the radiation environment at SFSIs and operational concerns regarding deployability as a safeguards instrument place significant constraints on detector requirements for dry cask imaging. Here we briefly comment on several possible muon detector technology choices as they pertain to cask monitoring.

Silicon-based tracking detectors are in use at most large collider experiments, and provide precision charged particle tracking with single hit position resolution better than 30 microns \cite{FVTXNIM, ATLAS, CMS, LHCb}.  While individual sensors typically have active areas of a few tens of cm$^{2}$, they can be arranged to cover tens of square meters in cylindrical or flat panel geometries; however, the channel count and subsequent cost of readout electronics is significant for large applications.  The high density electronics also require active liquid cooling in most cases, and transportation of a cooling system along with the detector in the field may not be feasible.  In addition, silicon detectors have well-known susceptibilities to radiation damage \cite{RadDamage1, RadDamage2}, and may deteriorate quickly in the significant neutron and gamma flux present at SFSIs. 

Plastic scintillators are a common charged particle detector with fast rise times and high efficiency.  Muon scattering tomography has been demonstrated with a tracker made of scintillating bars coupled to photomultiplier tubes, giving an intrinsic hit position resolution of 2.5 mm \cite{CRIPT}. Layers of scintillating fibers are also being explored for muon imaging nuclear waste storage vessels \cite{ClarksonDetector}.  In the radiation environment at a spent fuel storage installation, these detectors may suffer from  their sensitivities to both neutrons and gamma rays.  The exact rates that can be tolerated depend on the detector geometry, radiation levels around the cask, and electronics timing. Readout with silicon photomultipliers is becoming increasingly common, but these devices also suffer from radiation damage and are subject to dramatic changes in gain and dark rate noise with temperature fluctuations that may occur at outdoor sites \cite{SiPMRadDamage}.

Gas-based tracking detectors, such as wire chambers or drift tubes, have been used in large particle physics experiments for decades \cite{Charpak, DCbook}.  As much of the active volume is gas, these detectors have a relatively low density and are therefore comparatively insensitive to the neutral radiation from the cask, while still retaining high efficiency for charged particles. The single hit position resolution is typically around several hundred microns.  Several groups have used drift chambers\cite{Pesente} and drift tubes \cite{LANL_LMT, Russians} in the laboratory.  In practical terms, drift chambers with open volumes between anode wires are subject to electrical shorts when wires break and wrap around intact wires, which can be of concern when transporting chambers to and around SFSIs.  Drift tubes have individualized detector elements that can be arranged in planar or cylindrical geometries, eliminating interference between channels.  We note here that the only measurement of muon scattering radiography on an actual fuel cask successfully used drift tube detectors \cite{DurhamCask}.

\section{Summary}

For the first time, image reconstruction with cosmic ray muons using computed tomography algorithms has been studied.  GEANT4 simulations of muon imaging of a partially loaded dry storage cask show that missing fuel assemblies can be located with high confidence in less than two days, using a detector which completely surrounds the cask.  The diversion of spent fuel assemblies can therefore be determined without opening the cask, on a time scale that is well within the International Atomic Energy Agency timeliness goals.  Alternatively, a set of smaller detectors can be moved around the cask to generate tomographic information from multiple viewing angles.  This technique and a dedicated instrument could be a useful tool for international nuclear safeguards inspectors.

\begin{acknowledgments}
This work is funded by the National Nuclear Security Administration's Office of Defense Nuclear Nonproliferation Research \& Development. We thank Philip Winston of Idaho National Laboratory and Arden Dougan of NA-22 for useful discussions.
\end{acknowledgments}

\bibliography{LANL_cask_tomo}

\end{document}